*In memory of Prof. Ya. A. Smorodinsky*

# Hydrogen Atom in Electric and Magnetic Fields: Dynamical Symmetries, Superintegrable and Integrable Systems, Exact Solutions

## Mikhail A. Liberman

The Nordic Institute for Theoretical Physics (NORDITA), Royal Institute of Technology and Stockholm University. Hannes Alfvéns väg 12, SE-106 91 Stockholm, Sweden

## Abstract

The hydrogen atom is a supersymmetric system with the $SO(4)$ symmetry group generated by the angular momentum vector and the Runge-Lenz vector. The Schrödinger equation for the hydrogen atom and the corresponding Hamilton-Jacoby equation are separable in several different systems of coordinate, and the problems under consideration admit an exact analytical solution. The problem of a hydrogen atom in a uniform electric field (Stark effect) is integrable, it is separable in parabolic coordinates and possesses the symmetry group $SO(2) \times SO(2)$. The problem of the hydrogen atom in a uniform magnetic field is not separable. It is considered as an example of quantum chaos, that cannot be solved by any analytical method. Nevertheless, an exact analytical solution can be obtained in the form of a convergent power series in two variables, the radius and the sine of the polar angle. Therefore, the problem can be considered superintegrable, although it does not possess supersymmetry.

Highlights

- Hydrogen atom in a Magnetic Field
- Integrable and Superintegrable System
- No-separable Quantum Superintegrable System
- Exact analytical solution and dynamical symmetry

# 1. Introduction

The availability of an exact analytical solution to the physical problem is of great importance, providing an understanding of the phenomena in question over a wide range of parameters, as well as providing a solid fundamental platform on which further developments can be carried out. It is known that if a system is separable (Hamilton-Jacobi equation in classical mechanics or Schrödinger equation in quantum mechanics), it is integrable and, at least in principle, can be solved explicitly and sometime analytically. The integrability of a system is closely related to the separation of variables in a particular coordinate system, which means that there are at least N-1 integrals of motion, where N is the dimensionality of the system. Geometric symmetry means that the Hamiltonian of the system is invariant with respect to transformations in space (rotation, reflection, etc.), more generally, dynamical symmetry (also called "hidden" symmetry) implies that the Hamiltonian, matrix elements, etc. are invariant under transformations in the phase space of the corresponding symmetry group, which may also include the geometric symmetry group as a subgroup. Well-known examples of systems with dynamical symmetry are the Kepler problem in classical mechanics, the hydrogen atom and the harmonic oscillator in quantum mechanics, for which the Hamilton-Jacobi equation and the Schrödinger equation are separable in several different coordinate systems, which means that the number of integrals of motion is larger than N-1. The interest to the symmetry group in quantum mechanics has been partly grown due to the progress in classification of elementary particles according to the irreducible representations of various Lie groups, and became a new area dealing with symmetry, degeneracy, and accidental degeneracy. The relationship between symmetry and degeneracy of a system, that is prescribed by the dimensions of the irreducible representations of its symmetry group is well-known. Yet, there is no restriction preventing a higher multiplicity of degeneracy (it is called an accidental degeneracy) than that required by a geometrical symmetry.

The systematic study of integrable and superintegrable systems with dynamical symmetries possessing a complete set of integrals of motion and corresponding " accidental" degeneracies began with the pioneering works of Ya. A. Smorodinsky, P. Winternitz and their students in the mid-1960s [1-4], where a classification of all possible integrable and superintegrable three-dimensional systems with a scalar potential was created. It has also been proved that the presence of two first- or second-order in momentum integrals of motion guarantees the separation of the variables in the corresponding Hamilton-Jacoby or Schrödinger equation, and that the converse is also true [3]. It was first noted in [1, 4] that if a system is separable, then it is integrable with integrals of motion of at most quadratic in momentum. This is related to the fact that the method of separation of variables can be used in the classical Hamilton-Jacoby formalism to find the integrals of motion (see Ref. [5] § 48).

A classical mechanics system is called integrable if it admits three integrals of motion, including the Hamiltonian, which are mutually Poisson commutative and are functionally independent constants of motion. The extension of this definition to a system with $N$ degrees of freedom and to quantum mechanics is straightforward, with replacing the Poisson brackets with commutators, and the Hamiltonian and constants of motion with operators that form an algebraically independent set. The system is superintegrable if it allows more integrals of motion than the number of degrees of freedom, which commute with the Hamiltonian, but not necessarily with each other. The important property of superintegrable systems is that the Hamilton-Jacoby and Schrödinger equations are separable in more than one coordinate system, and each separable coordinate system is associated with commuting integrals (operators) of the motion. Later, the existence of superintegrable systems with integrals of motion, which are polynomials of the third and higher order in momentum was discussed in [6-17]. For superintegrable quantum systems the Schrödinger equation is separable in different coordinate systems. The spectrum of these systems shows higher or "accidental" degeneracies that do not follow from the geometrical symmetries of the problem. It was suggested that superintegrable systems are exactly solvable, although there is no rigorous meaning of the term exactly solvable. For example, it was suggested in [11] that the Schrödinger equation is exactly solvable if it is separable in some coordinate system and the solution of each resulting ordinary differential equation can be expressed in terms of hypergeometric functions. It is known that the first-order symmetry operators correspond to the explicit geometric symmetries, whereas the second-order operators correspond to "hidden" dynamical symmetries. The best-known example of a superintegrable system is the Coulomb potential, for which the Schrödinger equation is separable in four coordinate systems: spherical, parabolic, conical, and prolate spheroidal, which corresponds to 4 different pairs of independent components of the integrals of the angular momentum $\vec{M}$ and the Runge-Lenz vector $\vec{R}$.

After the first studies of integrable and superintegrable Hamiltonian systems with a scalar potential, the research in this area was extended to Hamiltonian systems with a vector potential. At present there is a large amount of literature devoted to the study of integrable and superintegrable Hamiltonian systems, in particular, in a magnetic field [18-32]. It was shown [23] that, in contrast to the case of a scalar potential, the existence of a second-order integral of motion in the case of a vector potential does not guarantee the separation of variables, although in the case of a two-dimensional problem the solution can be obtained using the concept of "quasi-separation of variables".

In the present work we use the hydrogen atom in external electromagnetic fields as classical examples of superintegrable, integrable, and nonintegrable systems, which are, respectively, the hydrogen atom without external fields, the hydrogen atom in a uniform electric field (Stark effect), and the hydrogen atom in a uniform magnetic field (Zeeman effect). Although the variables in the Schrödinger equation for a hydrogen atom in a magnetic field are not separable, and the system is traditionally considered as non-integrable according to the conventional classification, the solution for the hydrogen atom in a magnetic field can be obtained in the form of infinite converging power series in the radial variable with coefficients being polynomials in the sine of the polar angle. Because of the two-dimensionality, the wave functions are determined by solving an infinite system of equations, in contrast to the usual one-dimensional problems of mathematical physics, where eigenvalues are defined implicitly via a single equation. The solution can be viewed as an analytical function of two variables, similar to well-known analytical functions, like e.g. hypergeometric functions, which are defined as infinite converging series and are exact because their quantitative determination is free of any approximations. By analogy with the usual special functions of mathematical physics, the solution can be calculated algebraically with any desired accuracy [33, 34].

## 2. The Hydrogen atom and the Kepler problem
### 2.1 The Kepler problem

The best-known superintegrable systems with hidden symmetry are the Kepler problem and its quantum-mechanical counterpart, the hydrogen atom. A dynamic symmetry of the non-relativistic Kepler problem, which is higher than the space rotation symmetry, has been known for a long time. The most important consequence of the space and time symmetries is the existence of conservation laws, such as the conservation of energy, momentum, and angular momentum. For each global continuous symmetry, there is an integral of motion associated with it. This is the famous theorem, proved by Emmy Noether [35] in 1918, stating that for every differentiable symmetry of a physical system with conservative forces there is a corresponding conservation law. Dynamic symmetries are additional symmetries that are realized in phase space, and the Hamiltonian formulation is most convenient for their study.

The system with a central force is obviously rotationally symmetric with the conserved angular momentum, $\vec{M} = [\vec{r} \times \vec{p}]$. The constancy of $\vec{M}$ means that the radius vector of the particle $\vec{r}$ in a central field lies in the $(x, y)$ plane perpendicular to $\vec{M}$. The trajectory of a particle in the central potential field is limited by the plane $(x, y)$ perpendicular to the angular momentum, and in general case it is not closed and the classical motion is non-ergodic. Using the Lagrangian in polar co-ordinates $L = \frac{m}{2}(\dot{r}^2 + r^2\dot{\varphi}^2) - U(r)$, expressing $\dot{\varphi}$ in terms of the angular momentum, $p_\varphi = mr^2\dot{\varphi}$, where the generalized momentum $p_\varphi = M_z = M$, we obtain for the energy [5]:

$$E = \frac{\mu \dot{r}^2}{2} + \frac{M^2}{2\mu r^2} + U(r), \qquad (2.1)$$

where $\mu = \frac{mm_1}{m+m_1}$ is the reduced mass. In general case of a finite motion, the trajectory is not closed and densely fills the annulus bounded by the circles $r = r_{max}$ and $r = r_{min}$, which are defined by Eq. (2.1) with $\dot{r} = 0$.

The nonrelativistic Kepler problem is the motion of a particle in the inverse square attractive Newtonian central force, the gravity attractive potential $U(r) = -\gamma mm_1/r$. While the planarity of the orbit for the Kepler problem is a consequence of the conservation of angular momentum, the orbit in the Kepler problem is unique among all central force problems in that it is periodic motion along closed curve. This is due to the fact that the Kepler problem has an additional constant of motion - the Runge-Lenz vector

$$\vec{R} = [\vec{\dot{r}} \times \vec{M}] - \alpha \frac{\vec{r}}{r} \equiv \frac{1}{\mu}[\vec{p} \times \vec{M}] - \alpha \frac{\vec{r}}{r}. \qquad (2.2)$$

Although this vector is called Runge-Lenz vector it was discovered by none of them[1].

---

[1] According to Goldstein investigation [36], the priority for its discovery in the classical mechanics goes back to the early 18th century. The first it was mentioned in 1710 in letters of Jakob Hermann [37] and Johann Bernoulli [38]. Almost hundred years later, in 1799, it was rediscovered by Laplace [39]. Later, Hamilton [40] in 1847 derived the existence of a new constant of the motion for the Kepler problem, which he called the "eccentricity vector." Much later, this vector appeared in a popular German textbook on vectors by C. Runge [41], which was referenced by W. Lenz in his paper on the old quantum mechanical treatment of the Kepler problem or hydrogen atom [42]. Presumably, the name Runge-Lenz vector became widely used after famous work by W. Pauli [43].

The simplest way to show that the Runge-Lentz vector is the integral of motion is to multiply both sides of the equation of motion by the vector of angular momentum:

$$\vec{M} \times \frac{d\vec{p}}{dt} = \frac{d}{dt}[\vec{M} \times \vec{p}] = -\frac{\alpha\mu}{r^3}[\vec{r} \times [\frac{d\vec{r}}{dt} \times \vec{r}]] = -\alpha\mu \frac{d}{dt}\left(\frac{\vec{r}}{r}\right). \quad (2.3)$$

From Eq. (2.3) it is seen that $\frac{d}{dt}\left\{[\vec{p} \times \vec{M}] - \alpha\mu \frac{\vec{r}}{r}\right\} = 0$, and $\vec{R} = \frac{1}{\mu}[\vec{p} \times \vec{M}] - \alpha \frac{\vec{r}}{r} = $ const. Note, that the existence of the Runge-Lentz vector means only one additional integral of motion because there are two relations:

$$(\vec{M} \cdot \vec{R}) = \frac{1}{\mu}(\vec{M} \cdot [\vec{p} \times \vec{M}]) - \frac{\alpha}{r}([\vec{r} \times \vec{p}] \cdot \vec{r}) = 0, \quad (2.4)$$

$$R^2 = \frac{1}{\mu^2}[\vec{p} \times \vec{M}]^2 - 2\frac{\alpha}{\mu r}([\vec{r} \times \vec{p}] \cdot \vec{M}) + \alpha^2 = \frac{1}{\mu^2}[\vec{p} \times \vec{M}]^2 - 2\frac{\alpha}{\mu r}\vec{M}^2 + \alpha^2 =$$
$$= \frac{1}{\mu^2}\vec{p}^2\vec{M}^2 - 2\frac{\alpha}{\mu r}\vec{M}^2 + \alpha^2 = \frac{2}{\mu}M^2\left(\frac{p^2}{2\mu} - \frac{\alpha}{r}\right) + \alpha^2 = \frac{2}{\mu}M^2 H + \alpha^2 \quad (2.5)$$

Since relations (2.4) and (2.5) do not depend on the choice of orbit parameters, there are 4 independent integrals of motion. It can be shown (e.g., [5] §15) that Runge-Lenz is directed along the perihelion of the elliptic orbit, $\vec{R} = \alpha e \vec{r}_p / r_p$, which means that conservation of the Runge-Lentz vector fixes the perihelion of the elliptic orbit in $(x, y)$ plane. Note, that a small deviation of the potential from the Newtonian $U(r) = -\alpha/r$ causes a slow precession of the ellipse, and the orbit is filling a strip between two circles $r_{max} = a(1+e)$ and $r_{min} = a(1-e)$, where $e < 1$ is the eccentricity, $a$ is the semi-axis of the ellipse.

## 2.2 Poisson brackets and Symmetry group

The Lie algebra consists of the symmetry generators, which are the symmetry-associated invariants. The structure parallel to the Lie commutators is the Poisson brackets of the corresponding integrals of motion. A classic example is the $O(3)$ group of 3D space rotations with rotation generators formed by angular momentum:

$$\{M_i, M_j\} = \varepsilon_{inm} r_n \varepsilon_{jmk} p_k - \varepsilon_{imk} p_k \varepsilon_{jnm} r_n = \varepsilon_{ijl} M_l, \tag{2.6}$$

where $\varepsilon_{ijk}$ is the Levi-Civita tensor. The Poisson brackets $\{M_i, R_j\}$ are

$$\{M_i, R_j\} = \varepsilon_{inm} r_n \left[ \left( \vec{p}^2 \delta_{jm} - p_j p_m \right) + \frac{\alpha \mu}{r} \left( \frac{r_j r_m}{r^2} - \delta_{jm} \right) \right] -$$
$$- \varepsilon_{imk} p_k \left[ 2 r_j p_m - \delta_{jm} (r_k p_k) - p_j r_m \right] =$$
$$p^2 \varepsilon_{inj} r_n - p_j \varepsilon_{inm} r_n p_m + \frac{\alpha \mu}{r} \left( \frac{\varepsilon_{inm} r_n r_m r_j}{r^2} - \varepsilon_{inj} r_n \right) - \tag{2.7}$$
$$\left[ 2 r_j \varepsilon_{imk} p_m p_k - \varepsilon_{ijk} p_k (r_k p_k) - \varepsilon_{imk} p_j r_m p_k \right] =$$
$$- p^2 \varepsilon_{ijn} r_n + \frac{\alpha \mu}{r} \varepsilon_{ijn} r_n + \varepsilon_{ijn} p_n (\vec{r} \cdot \vec{p}) = \varepsilon_{ijn} \left( [\vec{p} \times \vec{M}]_n + \frac{\alpha \mu}{r} r_n \right) = \varepsilon_{ijn} R_n$$

For the Poisson brackets $\{R_i, R_j\}$ we obtain

$$\{R_i, R_j\} = -p^2 \delta_{ij} (\vec{r} \cdot \vec{p}) - p_i r_j p^2 + 2(\vec{r} \cdot \vec{p}) p_i p_j + \frac{\alpha \mu}{r} \left( \frac{r_i r_j (\vec{r} \cdot \vec{p})}{r^2} - 2 r_j p_i + \delta_{ij} (\vec{r} \cdot \vec{p}) \right) +$$
$$\vec{p}^2 \delta_{ij} (\vec{r} \cdot \vec{p}) + \vec{p}^2 p_j r_i - 2 p_i p_j (\vec{r} \cdot \vec{p}) - \frac{\alpha \mu}{r} \left[ \frac{r_j r_i (\vec{r} \cdot \vec{p})}{r^2} - 2 r_j p_i + \delta_{ij} (\vec{r} \cdot \vec{p}) \right] = \tag{2.8}$$
$$\vec{p}^2 (p_i r_j - p_j r_i) - \frac{2 \alpha \mu}{r} \left( r_i p_j - r_j p_i \right) = -2 \mu \left( \frac{\vec{p}^2}{2\mu} - \frac{\alpha}{r} \right) \varepsilon_{ijk} M_k = -2 \mu H \varepsilon_{ijk} M_k$$

Since for bound states $2\mu H = 2\mu E = const < 0$, we can redefine the Runge-Lentz vector as

$$\mathcal{R}_i = R_i / \sqrt{-2\mu H}. \tag{2.9}$$

Then the Poisson brackets for $M_i$ and $\mathcal{R}_i$ are

$$\{M_i, M_j\} = \varepsilon_{ijk} M_k, \tag{2.10a}$$

$$\{M_i, \mathcal{R}_j\} = \varepsilon_{ijk} \mathcal{R}_k, \tag{2.10b}$$

$$\{\mathcal{R}_i, \mathcal{R}_j\} = \varepsilon_{ijk} M_k \tag{2.10c}$$

The Poisson brackets for angular momentum $M_i$ represents the space rotation group $SO(3)$ inherent to motion at any central potential. Relations (2.10a,b,c) for six components of $M_i$ and $\mathcal{R}_j$ form $SO(4)$ group with subgroup $SO(3)$. Introducing $\vec{G}^{\pm} = \frac{1}{2}(\vec{M} \pm \vec{\mathcal{R}})$, equations (2.10a,b,c) can be written as

$$\{G_i^{\pm}, G_j^{\pm}\} = \varepsilon_{ijk} G_k^{\pm}, \tag{2.11}$$

$$\{G_i^{+}, G_j^{-}\} = 0. \tag{2.12}$$

Relations (2.11) – (2.12) show that the group $SO(4)$ formed by $\vec{G}^{\pm}$ is isomorphic to direct production of two $SO(3)$ groups: $SO(4) = SO(3) \times SO(3)$.

It should be noted Bertrand's theorem [44], which states that in the problems of classical mechanics with the Hamiltonian consisting of the sum of the kinetic energy and the spherically symmetric potential, only Coulomb (gravitational) potential ($U(\mathrm{r}) \propto 1/r$) and the harmonic oscillator ($U(\mathrm{r}) \propto r^2$) have bound and closed orbits.

## 2.3 The Hydrogen atom

The quantities considered in the previous section in classical mechanics should be replaced by corresponding quantum mechanical operators, which keep their meaning according to the correspondence principle. Quantum analogs of Poisson brackets are commutators $[\hat{A}, \hat{B}] = i\hbar \{A, B\}$, where $A, B$ are classical functions of $\vec{r}$ and $\vec{p}$ inside the Poisson brackets $\{,\}$, while $\hat{A}, \hat{B}$ are the corresponding operators inside the commutators $[,]$. Operators $\hat{p}_i, \hat{r}_j$ and $\hat{M}_i = \varepsilon_{ijk} \hat{r}_j \hat{p}_k$, are hermitian, but the operator $\hat{\vec{p}} \times \hat{\vec{M}}$ in Eq. (2.2) and therefore the classical form of the Runge-Lenz vector is not hermitian

$$\begin{aligned}(\hat{\vec{p}} \times \hat{\vec{M}})_i^{\dagger} &= \varepsilon_{ijk} \hat{M}_k \hat{p}_j = \varepsilon_{ijk} \hat{p}_j \hat{M}_k + \varepsilon_{ijk} i\hbar \varepsilon_{kjl} \hat{p}_l = \\ \varepsilon_{ijk} \hat{p}_j \hat{M}_k &- 2i\hbar \delta_{il} \hat{p}_l = \varepsilon_{ijk} \hat{p}_j \hat{M}_k - 2i\hbar \hat{p}_i = (\hat{\vec{p}} \times \hat{\vec{M}})_i - 2i\hbar \hat{p}_i\end{aligned} \tag{2.13}$$

$$(\hat{\vec{M}} \times \hat{\vec{p}})_i^{\dagger} = (\hat{\vec{M}} \times \hat{\vec{p}})_i + 2i\hbar \hat{p}_i \tag{2.14}$$

W. Pauli [43] symmetrized the Runge-Lenz vector, which is an observable quantity and therefore a Hermitian operator. From (2.13-14) it is obvious that the Hermitian form of the Runge-Lenz vector is

$$\hat{\vec{R}} = \frac{1}{2\mu}([\hat{\vec{p}} \times \hat{\vec{M}}] - [\hat{\vec{M}} \times \hat{\vec{p}}]) - \frac{\alpha \vec{r}}{r} \qquad (2.15)$$

By using relation $[\hat{A}, \hat{B}] = i\hbar\{A, B\}$ we obtain $[\hat{M}_i, \hat{H}] = 0$ and $[\hat{R}_i, \hat{H}] = 0$, and

$$[\hat{M}_i, \hat{M}_j] = i\hbar \varepsilon_{ijk} \hat{M}_k, \qquad (2.16a)$$

$$[\hat{R}_i, \hat{M}_j] = i\hbar \varepsilon_{ijk} \hat{M}_k, \qquad (2.16b)$$

$$[\hat{R}_i, \hat{R}_j] = -2i\hbar \mu H \varepsilon_{ijk} \hat{M}_k. \qquad (2.16c)$$

Here $\hat{H} = \frac{\hat{p}^2}{2\mu} - \frac{\alpha}{r}$, $\alpha = \frac{e^2}{4\pi\varepsilon_0}$, the reduced mass $\mu = m_e m_p / (m_e + m_p) \approx m_e$.

Quantum analogs of (2.4) and (2.5) are

$$\hat{\vec{R}} \cdot \hat{\vec{M}} = \hat{\vec{M}} \cdot \hat{\vec{R}} = 0, \qquad (2.17)$$

$$\left(\hat{\vec{R}}\right)^2 = 2\frac{\hat{H}}{\mu}(\hat{M}^2 + \hbar^2) + \alpha^2 \qquad (2.18)$$

In 1926, Schrödinger [45] and Pauli [43] almost simultaneously published papers in which the spectrum of the hydrogen atom was calculated by solving the Schrödinger equation in [45] and using the algebraic method in [43]. Since operators $\hat{G}_k^\pm$ commute according to (2.11) and (2.12), their algebras are decoupled, and each of them constitutes $SO(3)$ algebra, which means that the eigenvalues of $(\hat{\vec{G}}^+)^2$ and $(\hat{\vec{G}}^-)^2$ are

$$(\hat{\vec{G}}^+)^2 = i(i+1)\hbar^2, \quad i = 0, \tfrac{1}{2}, 1, ... \qquad (2.19)$$

$$(\hat{\vec{G}}^-)^2 = (\hat{\vec{G}}^+)^2 = k(k+1)\hbar^2, \quad k = 0, \tfrac{1}{2}, 1, ... \qquad (2.20)$$

Taking into account Eq. (2.5), and $\hat{\mathcal{R}}_i = \hat{R}_i / \sqrt{-2\mu E}$, we can write

$$(\hat{\vec{G}}^+)^2 = \frac{1}{4}\left(\hat{\vec{M}}^2 + \hat{\vec{\mathcal{R}}}^2\right) = -\frac{1}{4}\left(\frac{\mu\alpha^2}{2E} + \hbar^2\right). \tag{2.21}$$

Then, from Eq. (2.21) the energy is

$$E = -\frac{\mu\alpha^2}{2\left(4(\hat{\vec{G}}^+)^2 + \hbar^2\right)} = -\frac{\mu\alpha^2}{2\hbar^2\left(4k(k+1)+1\right)} = -\frac{Ry}{n^2}, \quad n = (2k+1) = 1, 2, \ldots \tag{2.22}$$

where the Rydberg unit of energy is: $Ry = \dfrac{m_e e^4}{16\pi^2 \varepsilon_0^2 \hbar^2}$, and $\mu \simeq m_e$.

The symmetry group $SO(4)$ of the hydrogen atom was established after the fundamental works by V. Fock [46, 47] and V. Bargman [48]. Fock [46, 47] reduced the Schrödinger equation for the hydrogen atom in momentum space to an integral equation for spherical harmonics in four variables and showed that in momentum space the integral Schrödinger equation becomes an eigenvalue equation for hyperspherical harmonics. V. Bargmann showed that this is closely related to Pauli's treatment of the hydrogen atom [43] and demonstrated the connection between Fock's analytical approach and Pauli's algebraic approach. The $SO(4)$ symmetry of the hydrogen atom explains the accidental degeneracy of the energy terms and superintegrability of the problem.

## 3  The Hydrogen atom in a constant electrical field
### 3.1  Stark effect

In a uniform electric field $\vec{E}$, the SO(4) symmetry and an accidental degeneracy inherent to the hydrogen atom are broken, and the splitting in the energy spectrum is known as Stark effect. It is known that the Hamilton-Jacobi and the Schrödinger equation for the hydrogen atom in the uniform electric field are separable in parabolic coordinates [5, 49-52]. This is due to the existence of two integrals of motion: the projection of angular momentum onto the z-axis along the electric field $\vec{E}$ and the projection on the z-axis of the generalized Runge-Lenz vector. The generalized Runge-Lenz vector for the Stark effect was obtained by Redmond [49].

We choose the z-axis along the electric field $\vec{E} = (0,0,E)$. Obviously, the system is invariant with respect to the rotation around the z-axis, and the projection of the angular momentum onto the z-axis is the integral of motion, $d(\vec{M} \cdot \vec{E})/dt = 0$. Intuitively, the generalized Runge-Lenz vector $\vec{\mathcal{R}}_E$ must include its projection on the z-axis. Therefore, the simplest way is to calculate the time derivation of $(\vec{R} \cdot \vec{E})$ using the equation of motion of the electron in the Coulomb field with the electric field $\vec{E} = (0,0,E)$

$$\frac{d\vec{p}}{dt} = -\frac{\alpha}{r^3}\vec{r} + e\vec{E} \tag{3.1}$$

The time derivation of $(\vec{R} \cdot \vec{E})$ is

$$\frac{d(\vec{R} \cdot \vec{E})}{dt} = \frac{1}{\mu}[\frac{d\vec{p}}{dt} \times \vec{M}] \cdot \vec{E} + \frac{1}{\mu}[\vec{p} \times \frac{d\vec{M}}{dt}] \cdot \vec{E} - \frac{\alpha}{\mu}\frac{\vec{p}}{r} \cdot \vec{E} + \frac{\alpha}{\mu}\frac{\vec{r}(\vec{p} \cdot \vec{r})}{r^3} \cdot \vec{E} =$$
$$= \frac{e}{\mu}(\vec{r} \cdot \vec{E})(\vec{p} \cdot \vec{E}) - \frac{e}{\mu}\vec{E}^2(\vec{p} \cdot \vec{r}) = -\frac{e}{2}\frac{d}{dt}([r \times E])^2 \tag{3.2}$$

From equation (3.2) it follows that

$$\mathcal{R}_E = \vec{R} \cdot \vec{E} - \frac{e}{2}([\vec{r} \times \vec{E}])^2 = const \tag{3.3}$$

is the integral of motion, which is the projection on $\vec{E}$ of the generalized Runge-Lenz vector

$$\vec{\mathcal{R}}_E = \vec{R} - \frac{e}{2}[[\vec{r} \times \vec{E}] \times \vec{r}] \tag{3.4}$$

The Hermitian form of $\mathcal{R}_E$ in quantum mechanics is

$$\hat{\mathcal{R}}_E = \frac{1}{2\mu}\left(([\hat{\vec{p}} \times \hat{\vec{M}}] - [\hat{\vec{M}} \times \hat{\vec{p}}]) \cdot \vec{E}\right) - \left(\frac{\alpha\vec{r}}{r} \cdot \vec{E}\right) - \frac{e}{2}([\vec{r} \times \vec{E}])^2. \tag{3.5}$$

The states of the hydrogen atom in a constant electric field are characterized by energy and two quantum numbers corresponding to eigenvalues of the projection of the angular momentum onto the z-axis and the operator $\hat{\mathcal{R}}_E$. The symmetry of the problem is the $O(2) \times O(2)$ symmetry group. Separating the Schrodinger equation in parabolic coordinates,

Hughes [50] has shown that the eigenstates for the Stark effect can be interpreted as a particular realization of irreducible representations of the $O(4)$ group $|n_1 n_2 m\rangle$ expressed in terms of the generalized Laguerre polynomials $L_{n_1+|m|}^{|m|}$ and $L_{n_2+|m|}^{|m|}$.

### 3.2 Ion of the hydrogen molecule

The problem of two Coulomb centers – an ion of a hydrogen molecule is another system, for which there are two conserved quantities: the projections of the generalized Runge-Lenz and angular momentum onto the z axis along the internuclear axis. The study of the motion of an electron in the field of two Coulomb centers began with Pauli's paper [53]. Since then, this problem has been considered by many researchers [54-63]. The separation of variables in the Hamilton-Jacobi equation in elliptic coordinates has been known to Euler [64] since 1760 (see also the problem after § 78 in Ref. [5]). According to the classification [14] this system is integrable. The additional integral of motion – the projection of the generalized Runge-Lenz vector on the internuclear axis means the existence of the dynamical (hidden) symmetry in addition to spatial rotational symmetry about the internuclear axis. This symmetry also results in the separation of variables in the Schrodinger equation in elliptic coordinates.

The integral of motion can be obtained in a way, similar to that used in Sec. 3.1. The equation of motion is

$$\frac{d\vec{p}}{dt} = -\frac{Z_1}{r_1^3}\vec{r}_1 - \frac{Z_2}{r_2^3}\vec{r}_2 = -\frac{Z_1\vec{r}}{r^3} - \frac{Z_2(\vec{r}-\vec{r}_{12})}{|(\vec{r}-\vec{r}_{12})|^3}. \tag{3.6}$$

For simplicity, consider two nuclei with the same positive electric charge, separated by the distance $|\vec{r}_{12}|$ in the coordinate system with $\vec{r}_1 = \vec{r}$, $\vec{r}_2 = \vec{r} - \vec{r}_{12}$, and take $\mu = e = Z_1 = Z_2 = 1$. Similar to calculation of $\mathscr{R}_E$ in §3.1, it is expected that the generalized Runge-Lentz vector includes the projection of the Runge-Lentz vector onto the z-axis along $\vec{r}_{12}$. We calculate the time derivation of $(\vec{R} \cdot \vec{r}_{12})$ using Eq. (3.6)

$$\frac{d(\vec{R}\cdot\vec{r}_{12})}{dt} = [\frac{d\vec{p}}{dt}\times\vec{M}]\cdot\vec{r}_{12} + [\vec{p}\times\frac{d\vec{M}}{dt}]\cdot\vec{r}_{12} - \frac{d}{dt}\left(\frac{\vec{r}}{r}\right)\cdot\vec{r}_{12} + \frac{d}{dt}\left(\frac{(\vec{r}-\vec{r}_{12})}{\sqrt{(\vec{r}-\vec{r}_{12})^2}}\cdot\vec{r}_{12}\right)$$

$$= -\frac{(\vec{r}\cdot\vec{r}_{12})(\vec{p}\cdot\vec{r})}{r^3} + \frac{\vec{r}^2(\vec{p}\cdot\vec{r}_{12})}{|(\vec{r}-\vec{r}_{12})|^3} + \frac{(\vec{r}\cdot\vec{r}_{12})(\vec{p}\cdot\vec{r}_{12})}{|(\vec{r}-\vec{r}_{12})|^3} - \frac{\vec{r}_{12}(\vec{p}\cdot\vec{r})}{|(\vec{r}-\vec{r}_{12})|^3} \quad (3.7)$$

$$-\frac{(\vec{p}\cdot\vec{r}_{12})}{|(\vec{r}-\vec{R}_{12})|} + \frac{[(\vec{r}\cdot\vec{p})-(\vec{p}\cdot\vec{r}_{12})][(\vec{r}\cdot\vec{r}_{12})-\vec{r}_{12}^2)]}{|(\vec{r}-\vec{r}_{12})|^3} = 2\left(\frac{\vec{r}^2(\vec{p}\cdot\vec{r}_{12})}{|(\vec{r}-\vec{r}_{12})|^3} - \frac{(\vec{r}\cdot\vec{r}_{12})(\vec{p}\cdot\vec{r})}{|(\vec{r}-\vec{r}_{12})|^3}\right)$$

Using the vector identity $(A\times B)\cdot(C\times D) = (A\cdot C)(B\cdot D) - (A\cdot D)(B\cdot C)$ it is easy to see that the right-hand side in Eq. (3.7) is the total time derivative

$$\frac{2}{(\vec{r}-\vec{r}_{12})^3}\{r^2(\vec{p}\cdot\vec{r}_{12}) - (\vec{r}\cdot\vec{r}_{12})(\vec{p}\cdot\vec{r})\} = \frac{2([\vec{r}\times\vec{p}]\cdot[\vec{r}\times\vec{r}_{12}])}{(\vec{r}-\vec{r}_{12})^3} = \frac{d\vec{M}^2}{dt}. \quad (3.8)$$

Thus, the projection onto $\vec{r}_{12}$ of the generalized Runge-Lentz vector is

$$\mathcal{R}_2 = \left([\vec{p}\times\vec{M}] - \frac{\vec{r}}{r} + \frac{(\vec{r}-\vec{r}_{12})}{|(\vec{r}-\vec{r}_{12})|}\right)\cdot\vec{r}_{12} - \vec{M}^2 = const. \quad (3.9)$$

The quantity $\mathcal{R}_2$ can be written as a projection of the generalized Runge-Lenz vector on the axis between the nuclei $Z_1$ and $Z_2$

$$\vec{\mathcal{R}}_2 \cdot \frac{\vec{r}_{12}}{|\vec{r}_{12}|} = \left[\left([\vec{p}\times\vec{M}] - \frac{Z_1\vec{r}}{r} + \frac{Z_2(\vec{r}-\vec{r}_{12})}{|(\vec{r}-\vec{r}_{12})|}\right) - \vec{M}^2\frac{\vec{r}_{12}}{|\vec{r}_{12}|}\right]\cdot\frac{\vec{r}_{12}}{|\vec{r}_{12}|}, \quad (3.10)$$

where the expression inside the square brackets can be viewed as the generalized Runge-Lenz vector for the ion of hydrogen molecule.

The integral of motion $\mathcal{R}_E$ can be obtained as a limiting case of the hydrogen molecule ion, when the second nucleus is moved to infinity with its electric charge increasing as $|\vec{r}_{12}|^2$, so that the two-center problem reduces to that of the motion of an electron in the Coulomb field of the first nucleus and uniform electric field of the second nucleus.

Krivchenkov and Liberman [59] obtained the generalized Runge-Lenz vector by separating the Schrödinger equation in elliptic coordinates, however the final formula for the generalized Runge-Lenz vector obtained in [59] was written with mistake, which was corrected by Kryukov and Oks [62]. They followed the first few steps from Krivchenkov-Liberman's paper [59], arriving at the same expression in the elliptical coordinates as they did, for the additional conserved quantity, and then using the relation between the elliptical coordinates and the Cartesian coordinates, obtained correct expression Eq. (3.10) for the generalized Runge-Lenz vector.

In elliptical coordinates

$$\xi = \frac{r_1 + r_2}{r_{12}}, \eta = \frac{r_1 - r_2}{r_{12}}, \varphi = \arctg \frac{y}{x}, \qquad (3.11)$$

Using atomic units $\hbar = \mu = e = 1$, and taking $Z_1 = Z_2 = 1$, the Schrödinger equation is

$$\left\{ \frac{4}{r_{12}^2(\xi^2 - \eta^2)} \left[ \frac{\partial}{\partial \xi}(\xi^2 - 1)\frac{\partial}{\partial \xi} + \frac{\partial}{\partial \eta}(1 - \eta^2)\frac{\partial}{\partial \eta} + \frac{\xi^2 - \eta^2}{(\xi^2 - 1)(1 - \eta^2)} \frac{\partial^2}{\partial \varphi^2} \right] + \right. \\ \left. + \frac{4}{r_{12}(\xi + \eta)} + \frac{4}{r_{12}(\xi - \eta)} + 2E \right\} \psi(\xi, \eta, \varphi) = 0 \qquad (3.12)$$

Taking the wavefunction in the form $\psi(\xi,\eta,\varphi) = F(\xi)G(\eta)e^{im\varphi}$, after separation of variables we obtain two one-dimensional equations for $F(\xi)$ and $G(\eta)$ with the corresponding separation constants $\Lambda_\xi$ and $\Lambda_\eta$. After elementary but tedious calculations, the equations for $F(\xi)$ and $G(\eta)$ can be reduced to the eigenvalue equation $\hat{\mathcal{R}}_2 \psi(\xi,\eta) = \mathcal{R}_2 \psi$, where $\mathcal{R}_2(\xi,\eta)$ is the integral of motion in elliptic coordinates, which after transformation from elliptic to Cartesian coordinates gives expression (3.9).

The existence of two conserved operators: $\hat{M}_z = (\vec{M} \cdot \vec{R}_{12})/R_{12}$ and $\hat{\mathcal{R}}_2$, and separability of Schrödinger equation means that the symmetry of the problem is larger than the symmetry group SO(2) corresponding to the two-dimensional geometrical rotation generated by $\hat{M}_z$.

The symmetry group SO(2)×SO(2) generated by operators $\hat{M}_z$ and $\hat{\mathcal{R}}_2$, can be viewed as a remaining subgroup of the group SO(4) for the pure Coulomb problem [58, 59].

The electron terms of a hydrogen ion molecule are characterized by two quantum numbers corresponding to the eigenvalues of the operators $\hat{M}_z$ and $\hat{\mathcal{R}}_2$, which explains the applicability of the Neumann-Wigner theorem [65], which states that only energy terms with different symmetry can intersect. In other words, since the Schrödinger equation in this case allows complete separation of variables in the elliptic coordinate system, different energy terms are characterized by two quantum numbers corresponding to two separation constants [66]. Thus, the system is integrable and bound states of an electron in the field of two Coulomb centers admit, in some cases, simple analytic solutions [60, 61].

## 4. The Hydrogen atom in a uniform magnetic field

The systems considered in previous sections: a superintegrable hydrogen atom and an integrable hydrogen atom in a uniform electric filed exhibit supersymmetric features, in other words, the Schrödinger equation is separable in at least one coordinate system, and, as a rule, an exact analytical solution can be obtained. Recall that minimal integrability according to [14] means that the total number of functionally independent and mutually commuting integrals of motion (including the Hamiltonian) is equal to the number of degrees of freedom, which actually means that the Schrödinger equation is separable [67].

The hydrogen atom in a magnetic field is of great interest due to its important applications in astrophysics [68-70] and solid-state physics [71-74]. However, unlike to previously considered problems, in the presence of a uniform magnetic field the supersymmetry of the Coulomb problem is completely destroyed: there is neither a complete set of the constants of motion or ''good'' quantum numbers, except for the magnetic quantum number $m$ and the z-parity $\nu$ - reflection in the $x, y$-plane, which is perpendicular to the direction of the magnetic field, and the corresponding Schrödinger equation is not separable. For weak magnetic fields, $\mu_B B \ll Ry$, where $\mu_B$ is the Bohr magneton, the linear in $B$ term can be considered as a perturbation, and the corresponding solution for splitting of the energy levels of a hydrogen atom $\Delta E_n \propto mB$ (the Zeeman effect) is well known [66]. However, when the magnetic field strength increases and becomes comparable to the Coulomb field strength (the quadratic Zeeman regime), the problem totally loses both its spherical symmetry at $B \to 0$ and cylindrical symmetry at $\hbar\omega_c \gg e^2/r$, which is called the Landau regime with energy of the Landau levels $E_{Landau} = (N+1/2)B$, $N = 0,1,2,....$. In the intermediate region of the magnetic field strength any regularities of the energy spectrum are lost, and the hydrogen atom in a uniform magnetic field is considered as an example of quantum chaos. Since the corresponding Schrödinger equation is not separable, it was widely believed that the problem could not be solved by any analytical method.

The characteristic strength of the magnetic field can be obtained by equating the radius of the electron orbit for the lowest Landau level $r_0 = (\hbar c/eB)^{1/2}$ and the Bohr radius $r_B = \hbar^2/m_e e^2$, or equating the distance between Landau levels $\hbar\omega_B$ and the characteristic binding energy of the electron, $2Ry = m_e e^4/\hbar^2$, where $\omega_B = eB/m_e c$ is the cyclotron frequency, $B$ is the magnetic field strength. The characteristic magnitude of the magnetic field in quantum mechanics is $B_0 = \dfrac{m_e^2 e^3 c}{\hbar^3} = 2.35 \cdot 10^9 \, G$. Relativistic effects become noticeable for $\mu_B B > m_e c^2$, so that the nonrelativistic quantum mechanics is applicable for $B < m_e^2 c^3/e\hbar = 4.4 \times 10^{13} \, G$. Strong magnetic fields on the order of $B_0$ are far beyond to those available in laboratories, but for hydrogen-like (Wannier–Mott) excitons in semiconductors, due to the small effective mass $m_{eff} \ll m_e$ and large dielectric constants $\varepsilon \gg 1$ of semiconducting materials, $B_{eff} = m_{eff}^2 e^3 c / \hbar^3 \varepsilon^2 \ll B_0$. For example, for germanium $B_{eff} \approx 9 kG$, while the magnetic field for InSb can be considered strong already for $B > B_{eff} \approx 2 kG$, which means that the magnetic field in the range of 1T is already ultrahigh for these semiconductors.

As a result of intensive studies of integrable and superintegrable systems with vector potential (nonzero magnetic field) [18-32], for which the Schrödinger equation admits separation of variables in at least one coordinate system, the classification of such systems with corresponding pairs of operators commuting with the Hamiltonian has been obtained. Studies of integrable and superintegrable systems with a magnetic field [18–32] showed that the relation between separability and integrability is different compared to the case of scalar potential. It was noted [21, 23] that although the existence of integrals of the first and second order implies separation of variables in the Schrödinger equations for a scalar potential, this is not the case for $B \neq 0$. Recall that the first-order in momentum symmetry operators correspond to the explicit geometric symmetry of the problem, and the second-order operators correspond to the so-called "hidden" symmetries.

The existence of an exact analytical solution can usually indicate the hidden symmetry and superintegrability of the system and high-order integrals of motion. We will call the system exactly solvable if the solution can be obtained in the form of polynomials, which are defined as infinite convergent series with known recurrence relations for the coefficients and are exact, since their quantitative definition is free from any approximations, and the wave functions can be found with any precision as solutions of algebraic equations. However, although a hydrogen atom in a uniform magnetic field admits an exact analytical solution, as shown in Refs [33, 34], neither high-order integrals nor the "hidden" symmetry of the problem are known.

The Hamiltonian describing the motion of an electron in the Coulomb field of the nucleus and a homogeneous magnetic field $\vec{B}$ has the form

$$\hat{H} = -\frac{1}{2m_e}\left(\hat{\vec{p}} + \frac{|e|}{c}\hat{\vec{A}}\right)^2 - \frac{e^2}{r}. \tag{4.1}$$

We assume that the nucleus is infinitely heavy, $m_e / M_p \ll 1$, introduce a spherical system of coordinates $(r,\theta,\varphi)$ and a vector potential in the form $\vec{A} = [\vec{B}\times\vec{r}]/2 = (0,0,Br\sin\theta/2)$, where $\vec{B} = (B\cos\theta, -B\sin\theta, 0)$. Then, the Hamiltonian in spherical coordinates will be

$$\hat{H} = -\frac{\hbar^2}{2m_e}\Delta - i\hbar\frac{eB}{2m_e c}\frac{\partial}{\partial\varphi} + \frac{e^2 B^2}{8m_e c^2}r^2\sin^2\theta - \frac{e^2}{r}, \tag{4.2}$$

Using atomic units, $m_e = \hbar = c = 1$, and the dimensionless magnetic field $\gamma = B/B_0$, the Schrödinger equation $\hat{H}\Psi = E\Psi$ becomes:

$$\Psi_{rr} + \frac{2}{r}\Psi_r + \frac{1}{r^2}\Psi_{\theta\theta} + \frac{\cos\theta}{r^2\sin\theta}\Psi_\theta + \frac{1}{r^2\sin^2\theta}\Psi_{\varphi\varphi} +$$
$$+i\gamma\Psi_\varphi - \frac{1}{4}\gamma^2 r^2\sin^2\theta\,\Psi + \frac{2}{r}\Psi + 2E\Psi = 0 \tag{4.3}$$

where subscripts $r,\theta,\varphi$ denote corresponding partial derivatives.

For the wave function in the form

$$\Psi(r,\theta,\varphi) = e^{im\varphi}(r\sin\theta)^{|m|}(r\cos\theta)^\nu \psi(r,\theta), \tag{4.4}$$

the Schrödinger equation takes the form

$$\psi_{rr} + \frac{2(|m|+\nu+1)}{r}\psi_r + \frac{1}{r^2}\psi_{\theta\theta} + \frac{1}{r^2}\left[(2|m|+1)\cot\theta - 2\nu\tan\theta\right]\psi_\theta =$$
$$= \left[\frac{1}{4}\gamma^2 r^2\sin^2\theta - \frac{2}{r} - (1+|m|)\gamma + 2E_b\right]\psi \tag{4.5}$$

In Eq. (4.5) the total energy is expressed as $E = \gamma(|m|+m+1)/2 - E_b$, which coincides with the binding energy of the electron $\mathcal{E} = \gamma/2 - E$ for $m < 0$. We look for the solution of Eq. (4.5) in the form of a power series in $r$ with coefficients $f_i(t)$, where $t = \sin\theta$

$$\psi(r,t) = \sum_{i=0}^{\infty} f_i(t)r^i. \tag{4.6}$$

Substituting expansion (4.6) into Eq. (4.5) and assuming $f_k(t) \equiv 0$ for negative values of $k$, leads to the following equation for the coefficients $f_i(t)$:

$$(1-t^2)f_i'' + \frac{2|m|+1}{t} f_i' - 2(|m|+\nu+1)tf_i' + i(i+2|m|+2\nu+1)f_i = \qquad (4.7)$$
$$= \frac{1}{4}\gamma^2 t^2 f_{i-4} + [2E_b - \gamma(|m|+1)f_{i-2} - 2f_{i-1}$$

Equation (4.7) is an inhomogeneous linear differential equation, the solution of which is the sum of the solution of the corresponding homogeneous equation $F_i(t)$ and the partial integral $G_i(t)$ of equation (4.7). Substituting in the homogeneous part of Eq. (4.7) solution in the form $F_i(t) = \sum_{j=0}^{\infty} b_{i,j} t^j$ and equating the coefficients with the same powers $j$, we obtain recurrent relations for the coefficients $b_{i,j}$, which independently couple the coefficients with even and odd $j$:

$$b_{i,j+2} = -\frac{(i-j)[i+j+2(|m|+\nu)+1]}{(j+2)(j+2|m|+2)} b_{i,j}. \qquad (4.8)$$

Since all coefficients in the corresponding even or odd subset are uniquely determined by one coefficient in each subset, the solution of the homogeneous equation can be taken as a linear combination of two basis vectors. Therefore, any solution of the homogeneous equation may be taken as a linear combination of two basis vectors: the first vector corresponds to $b_{i,0}=1$, $b_{i,1}=0$, and the second one to $b_{i,0}=0$, $b_{i,1}=1$. The ratio $b_{i,j+2}/b_{i,j} \to 1$ for $j \to \infty$, meaning that at $t=1$ the function $F_i(t) \to \infty$, which does not happen if the series $F_i(t)$ terminates at a finite value of $j$. As it is seen from Eq. (4.8) this occurs if $i=j$. Therefore, even values of $i$ a physically allowable solution of the homogeneous equation involves only the first basis vector $b_{i,0}=1$, $b_{i,1}=0$, while for odd values of $i$ only the second basis vector $b_{i,0}=0$, $b_{i,1}=1$ is acceptable. In both cases the function $F_i(t)$ is the product of an arbitrary constant $C_i$ and a polynomial $H_i(t) = \sum_{j=0}^{i} h_{i,j} t^j$ with the lowest term equal to unity, and $b_{i,j} = C_i h_{i,j}$. Thus, the solution to Eq. (4.7) has the form

$$f_i(t) = G_i + C_i H_i(t). \tag{4.9}$$

It can be proved [33] that for any physically admissible solutions the particular integral is also a polynomial of degree $i$ in the form $G_i(t) = \sum_{j=0}^{i} a_{i,j} t^j$, and for a physically admissible solution the coefficients are given by the recurrent relation

$$a_{i,j+2} = \frac{(j-i)[i+j+2(|m|+\nu)+1]}{(j+2)(j+2|m|+2)} a_{i,j}. \tag{4.10}$$

Finally, the solution of Eq. (4.5) can be written as

$$\psi(r,\theta) = \sum_{i=0}^{\infty} \sum_{j=0}^{i} A_{i,j} r^i \sin^j \theta, \quad A_{i,j} = a_{i,j} + C_i h_{i,j}. \tag{4.11}$$

The function $\psi(r,\theta)$ must obey the boundary condition on the axis at $\theta = 0$:

$$\left. \frac{\partial \psi(r,\theta)}{\partial \theta} \right|_{\theta=0} = 0. \tag{4.12}$$

This means that $a_{i,j}$ and $b_{i,j}$ with odd $j$ must be zeros, consequently, $A_{i,j} = 0$, and the function $\psi(r,\theta)$ contains only even powers of $\sin\theta$.

We came to the solution which structure is in the form of a power series in two variables $r$ and $\sin\theta$:

$$\psi(r,\theta) = \sum_{i=0}^{\infty} \sin^{2k}\theta \sum_{j=0}^{i} A_{i,2k} r^i, \quad A_{i,2k} = \begin{cases} a_{i,2k} + C_i h_{i,2k}, & i = 2p \\ a_{i,2k}, & i = 2p+1 \end{cases}. \tag{4.13}$$

The polynomials $H_i(t)$ are nonzero only for even $i$

$$H_i(t) = \sum_{k=0}^{i/2} h_{i,2k} t^{2k}, \tag{4.14}$$

where $h_{i,0} = 1$, and $h_{i,2k}$ is defined by Eq. (4.8) for $j = 2k$

$$h_{i,2k+2} = -\frac{(i-2k)[i+2k+2(|m|+\nu)+1]}{(2k+2)(2k+2|m|+2)} h_{i,2k}. \tag{4.15}$$

If $\psi(0) \neq 0$ then in Eq. (4.13) one can take $C_0 = 1$; the remaining coefficients $C_{2p}$ and the eigenvalue $E_b$ are determined from the boundary condition at infinity, $\psi(r \to \infty, \theta) = 0$, that is reduced to a single nonlinear equation.

The use of high-precision floating-point arithmetic written as a portable C++ code (see [33]) allows efficient and fast calculations of sums in Eqs. (4.11) - (4.14). For example, the time required to obtain seven significant digits of the ground-state binding energy for the quadratic Zeeman effect ($\gamma = 1$) using a PC laptop, is about 1 sec, and for very strong magnetic field, $\gamma = 1000$, similar calculations take about one minute. The precision required to compute the ground-state binding energies with accuracy $10^{-12}$ Hartree is 38 decimal digits for $\gamma = 1$ and 280 decimal digits for $\gamma = 100$. The binding energies of the ground state $1s_0$, exciting stage states $2s_0, 2p_0, 2p_{-1}, 3p_0, 3p_{-1}, 3d_{-1}, 3d_{-2}$ and the energy levels of the lowest states with $\nu = 0$ and $m = 0, -1, -2, -3, ... -10$ were calculated in [33] with accuracy $10^{-12}$ Hartree for magnetic fields in the range $0 \leq \gamma \leq 4000$. The calculated electron density of the ground state $1s_0$ in a magnetic field of different strength $0 \leq \gamma \leq 1000$ are shown in Figure 1.

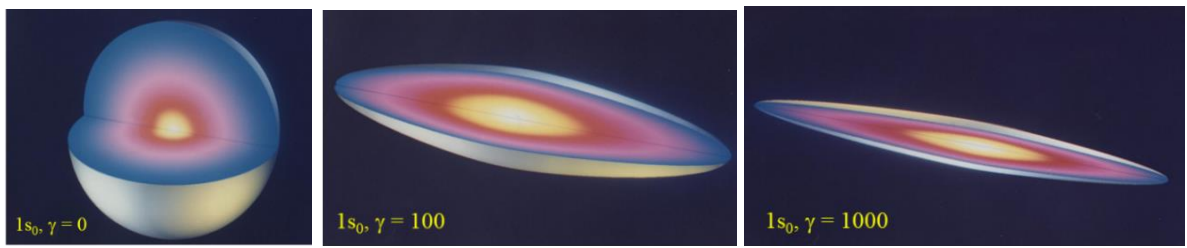

Figure 1. The $1s_0$ state of the hydrogen atom in magnetic fields: $\gamma = 0; 100; 1000$.

The calculated electron density for several excited states of the hydrogen atom and electron densities for the zeros Landau level are shown in Figure 2.

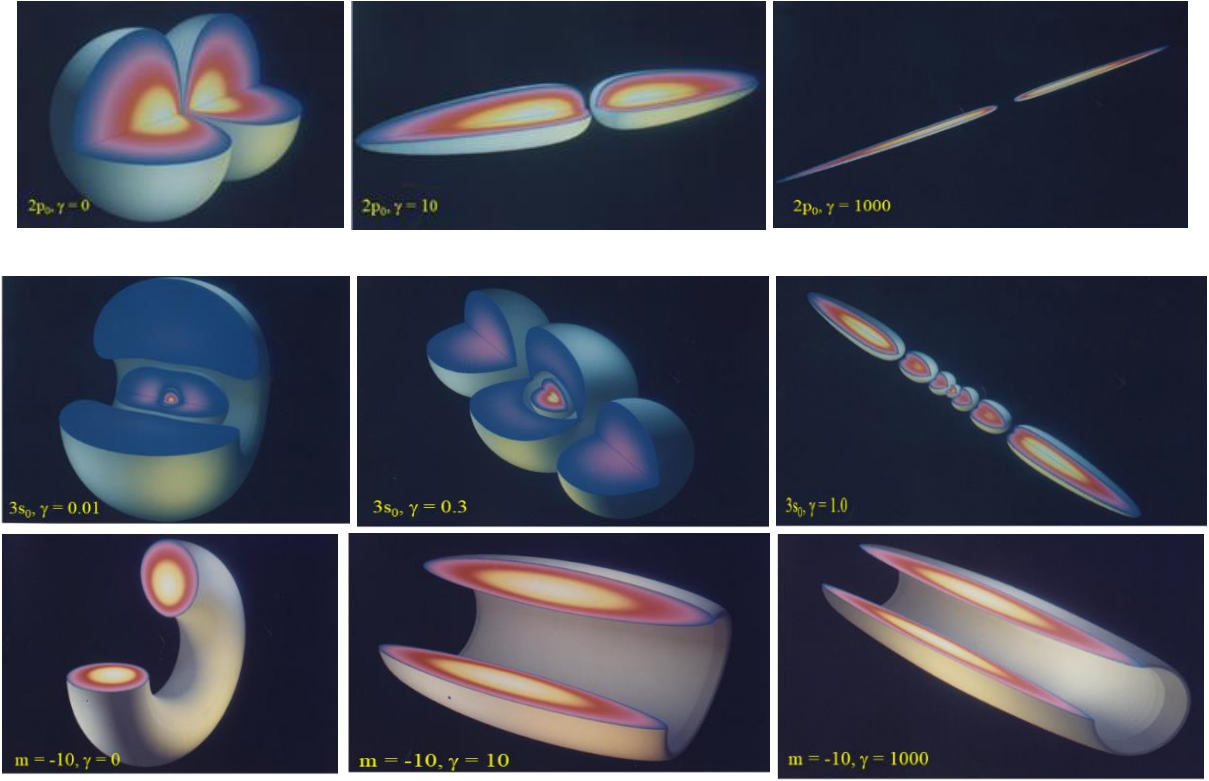

Figure 2. Exited states $2p_0$, $3s_0$ of the hydrogen atom and electron densities for the zeros Landau level (the lowest energy levels with quantum numbers the zeros parity and $m=-10$) in the magnetic fields of different strength. The magnitude of $\gamma$ is shown in figures.

The exact solution for a hydrogen atom in a uniform magnetic field helps to understand the transition from the ordered energy levels of a field-free hydrogen atom to a chaotic regime, when the symmetry breaks down with the increase of the magnetic field strength [75]. Since the gap between neighboring energy levels in a pure hydrogen atom decreases as $1/n^3$, the quadratic Zeeman effect (term $\propto \gamma^2$ in Eq. (4.3)) has the greatest effect at large values of $n$. For example, n-mixing occurs for the principal quantum numbers $n=6,7$ for $\gamma \approx 4.2 \cdot 10^{-3}$, and for $n=9,10$ it occurs for $\gamma \approx 9.22 \cdot 10^{-6}$. Fig. 3(a, b) shows the energy terms with $m=0$, $\nu=+1$ evolving from pure hydrogen atom states with principal quantum numbers $n=6,7,8,9,10$. The anticrossing shown in Fig. 3b is due to the existence of an approximate constant of motion. The width of the first avoided crossings between energy levels with principal quantum numbers $n=6$ and $n=7$ decreases exponentially when $n$ increases.

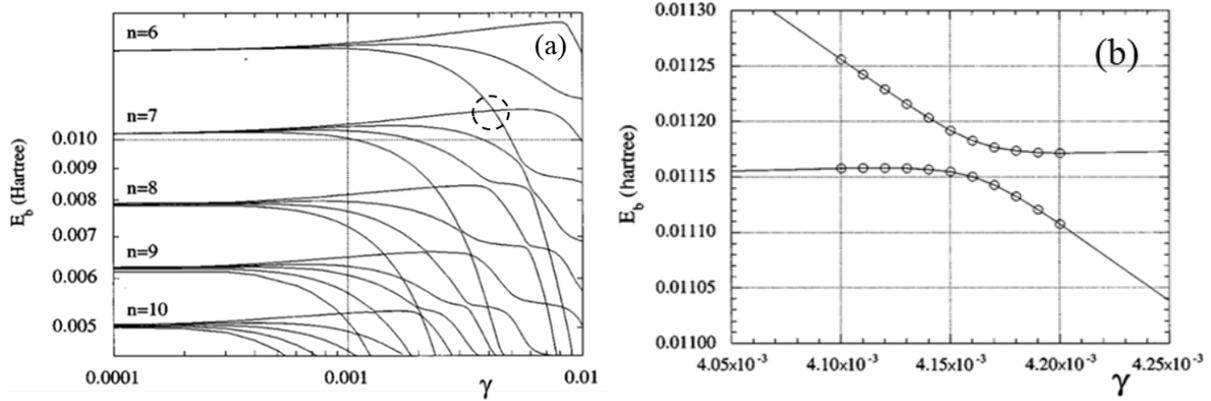

Figure 3(a, b). (a): Irregular behavior of the energy levels $m=0$, $\nu=+1$ evolving from the field-free states with principal quantum numbers $n=6,7,8,9,10$. (b): The avoided crossing between levels with principal quantum numbers $n=6$ and $n=7$ shown by dashed line circle in Fig. 3(a).

## 5. Conclusion

The field-free hydrogen atom is an example of a super-symmetric and super-integrable system, which has more integrals of motion than the number of degrees of freedom. Therefore, the Schrödinger equation are separable in several different coordinate systems, and the problem has an exact analytical solution. A uniform electric field destroys the high $SO(4)$ symmetry of the field-free hydrogen atom but leaves the lower $SO(2) \times SO(2)$ symmetry group, which is a remaining subgroup of the $SO_4$ symmetry group inherent to the field-free hydrogen atom. Accordingly, two integrals of motion remain, which are projections of the angular momentum and of the generalized Runge-Lentz vector on the electrical field direction. The Schrödinger equation in this case is separable in parabolic coordinates and therefore the system is considered as integrable.

A uniform magnetic field completely destroys the high, dynamic symmetry of the field-free hydrogen atom. In this case, only two "good" quantum numbers remain: the magnetic quantum number $m$ corresponding to rotation around the direction of the magnetic field and the z-parity $\nu$ corresponding to reflection in the plane perpendicular to the direction of the magnetic field. Although the Schrödinger equation in this case is not separable, an exact analytical solution can be obtained in the form of a convergent power series in two variables, the radius and the sine of the polar angle, with the boundary condition in the form of zero-dimensional conditions of the converging series of reduced solutions. Due to the two-dimensionality of the problem the wave functions are determined by solving an infinite system of equations, unlike the usual one-dimensional problems of mathematical physics, where eigenvalues are defined implicitly through a single equation. The obtained solution is similar to well-known analytic functions, such as, for example, the hypergeometric function, which is defined as an infinite convergent series and is exact, since its quantitative definition is free from any approximations.

According to classification of a super-symmetric, super-integrable system, such systems possess integrals of motion commuting with the Hamiltonian and the corresponding Schrödinger equation is separable. It is believed that superintegrable systems, at least in the case of a scalar potential, have an exact analytical solution and that the converse statement is true: a system that has an exact analytical solution is superintegrable, which means that it has more integrals of motion than the number of degrees of freedom. It seems that the case of a hydrogen atom in a uniform magnetic field apparently excludes the converse statement: the existence of an exact analytical solution does not guarantee that the system has additional integrals of motion and hidden symmetry. At least, all known studies of superintegrable systems in a magnetic field [18-32] did not find additional integrals of motion for this problem, and the only known integral is the projection of angular momentum.

## Acknowledgements
The author wishes to thank Alexander Zhalij, Alexey Magazev and Antonella Marchesiello for discussions and valuable comments.